%% file: main.tex
\documentclass[conference]{IEEEtran}

\usepackage{amsmath,amssymb,amsfonts}
\usepackage{graphicx}
\usepackage{textcomp}
\usepackage{xcolor}
\usepackage{booktabs} 

\usepackage{adjustbox}
\usepackage{hyperref}
\usepackage{subcaption}
\usepackage{algorithm}
\usepackage{algpseudocode}
\usepackage{amsmath}
\usepackage[inline]{enumitem} 

\def\BibTeX{{\rm B\kern-.05em{\sc i\kern-.025em b}\kern-.08em
    T\kern-.1667em\lower.7ex\hbox{E}\kern-.125emX}}
\usepackage{cite}
\usepackage{url}

\usepackage{etoolbox}           
\newcounter{observation}
\newenvironment{observation}[1][]{%
  \refstepcounter{observation}
  \noindent\textbf{Observation~\theobservation}%
  \ifblank{#1}{}{~(\emph{#1})}.%
  \enspace\ignorespaces            
}{\par\ignorespacesafterend}       

\begin{document}

\title{Performance Evaluation of Brokerless Messaging Libraries}


\author{
    \IEEEauthorblockN{Lorenzo La Corte}
    \IEEEauthorblockA{\textit{Hitachi Energy Research} \\
    Baden-Dättwil, Switzerland
    }
    \and
    \IEEEauthorblockN{Syed Aftab Rashid}
    \IEEEauthorblockA{\textit{Hitachi Energy Research} \\
    Baden-Dättwil, Switzerland 
    }
    \and
    \IEEEauthorblockN{Andrei-Marian Dan}
    \IEEEauthorblockA{\textit{Hitachi Energy Research} \\
    Baden-Dättwil, Switzerland
    }
}

\maketitle

\begin{abstract}
Messaging systems are essential for efficiently transferring large volumes of data, ensuring rapid response times and high-throughput communication.
The state-of-the-art on messaging systems mainly focuses on the performance evaluation of \emph{brokered} messaging systems, which use an intermediate broker to guarantee reliability and quality of service. 
However, over the past decade, \emph{brokerless} messaging systems have emerged, eliminating the single point of failure and trading off reliability guarantees for higher performance. Still, the state-of-the-art on evaluating the performance of brokerless systems is scarce.
In this work, we solely focus on brokerless messaging systems. First, we perform a qualitative analysis of several possible candidates, to find the most promising ones. 
We then design and implement an extensive open-source benchmarking suite to systematically and fairly evaluate the performance of the chosen libraries, namely, ZeroMQ, NanoMsg, and NanoMsg-Next-Generation (NNG). We evaluate these libraries considering different metrics and workload conditions, and provide useful insights into their limitations.
Our analysis enables practitioners to select the most suitable library for their requirements. 
\end{abstract}



\input{introduction}

\input{overview}

\input{qualitative}

\input{methodology}

\input{evaluation}

\input{related}

\input{conclusion}



\bibliographystyle{IEEEtran}
\bibliography{bib}

\end{document}

%% file: introduction.tex
\section{Introduction}\label{sec:introduction}

Efficiently transferring large amounts of data while ensuring quick response times is a significant challenge in control systems, Internet-of-Things (IoT), robotics, computer networks, etc. 
\emph{Messaging libraries} enable the exchange of messages/packets between various software components, offering high throughput and low latency.

Messaging libraries or frameworks typically use message queues to handle \emph{asynchronous} communication, ensuring reliable message transfer even when some components are temporarily unavailable. Additionally, most systems include a \emph{broker}, an intermediary service that routes and distributes messages between producers and consumers. Brokered systems usually offer features such as message persistence, guaranteed delivery, and built-in load balancing.

\emph{Brokerless} architectures eliminate the broker, removing the single point of failure and enabling publishers and subscribers to communicate directly, thus achieving higher performance in terms of latency~\cite{dworak2012new} and bandwidth~\cite{afanasev2017performance}.
However, this shifts responsibilities for reliability, discovery, and routing to the application layer, which in turn offers greater flexibility by allowing developers to tailor these aspects precisely to their requirements.
Moreover, brokerless libraries are typically lightweight, with a small memory footprint and minimal dependencies, making them particularly suited for resource-constrained environments~\cite{barroso2016benchmarking, patro2017comparative}. 
Therefore, brokerless frameworks have emerged in various contexts, such as control and monitoring systems~\cite{patro2017comparative}, IoT~\cite{bertrand2021classification}, and robotics~\cite{goertzel2014software, andreev2015technology, kirill2015software}.

Several state-of-the-art studies focus on evaluating the performance of brokered systems~\cite{sommer2018message, de2019performance, fu2020fair, bertrand2021classification, maharjan2023benchmarking}, considering performance metrics such as throughput, latency, CPU, and memory consumption.
However, to the best of our knowledge, only a handful of works in the state-of-the-art include brokerless solutions in their analysis~\cite{dworak2012new, barroso2016benchmarking, afanasev2017performance}, and most of their results are outdated considering the rapidly evolving landscape of messaging systems. 
Therefore, new insights on the performance capabilities and limitations of brokerless messaging frameworks are necessary.

In this work, we address the gap in the state-of-the-art 
by analyzing several brokerless libraries qualitatively and quantitatively.
Specifically, this work makes the following contributions:
\begin{enumerate}
    \item A qualitative analysis of existing brokerless messaging libraries, focusing on licensing, ease of installation, documentation, maintenance, and community support. Based on the qualitative analysis, we choose three libraries, namely ZeroMQ, NanoMsg, and NNG, for performance evaluations.
    \item An open-source benchmarking suite~\cite{suite} to systematically and fairly evaluate the performance of the three chosen libraries under different settings. The benchmarking suite can be easily extended to include other libraries.
    \item A thorough experimental evaluation to compare the performance of the chosen libraries, focusing on latency, throughput, jitter, CPU usage, and memory consumption. The results obtained show the trade-offs between the libraries in various scenarios, offering insights into their optimal use cases.
\end{enumerate}

The article is organized as follows. Section~\ref{sec:bml} outlines the general characteristics of messaging libraries. Section~\ref{sec:qualitative_analysis} presents a qualitative analysis of several brokerless messaging frameworks. Section~\ref{sec:methodology} describes our benchmarking suite, designed to fairly and systematically compare the performance of the selected libraries. Section~\ref{sec:performance_comparison} discusses the experimental results and the insights derived from them. Finally, Section~\ref{sec:related} reviews related work and Section~\ref{sec:conclude} concludes the article.

%% file: overview.tex
\section{Overview of Messaging Libraries}\label{sec:bml}

In this section, we discuss communication transports and patterns commonly offered by messaging libraries. Then, we proceed by highlighting key differences between brokered and brokerless architectures.

Messaging libraries typically rely on three transport mechanisms: 
\begin{enumerate*}
    \item the \emph{Transmission Control Protocol} (TCP) is primarily used for network communications and leverages the operating system's network stack to ensure reliable data transfer.
    \item \emph{Inter‑Process communication} (IPC) mechanisms, such as Unix domain sockets, allow two processes on the same machine to exchange data without traversing the network stack, through file‑system paths or unique local identifiers, generally incurring into lower latencies compared to the TCP communication.
    \item \emph{In‑Process communication} enables threads in the same process to exchange data directly via shared memory, and is therefore the fastest option, as it avoids the overhead of the operating‑system networking stack, significantly reducing communication latencies.
\end{enumerate*}

On top of the transport, messaging frameworks offer standardized communication patterns.
\emph{Request/Reply} (REQ/REP) is a message‑centric pattern, used for Remote-Procedure-Call (RPC)‑like interactions, where a message is sent as a request, with a corresponding reply expected from the receiver.
\emph{Publish/Subscribe} (PUB/SUB) is a data‑centric pattern, where publishers broadcast messages to interested subscribers, registered to specific topics. This provides space, time, and synchronization decoupling~\cite{eugster2003many}.


Traditional Message‑Oriented Middleware (MOM) platforms --- such as Apache Kafka\cite{kreps2011kafka} and RabbitMQ\cite{williams2012rabbitmq} --- use a \emph{message broker}, a long‑running server responsible for topic management, persistence, and routing.
The broker decouples producers and consumers in \emph{space} (they need not know each other) and \emph{time} (messages can be stored until a consumer reconnects), enabling different quality of service guarantees. These features come at the price of (i) an extra network hop (store‑and‑forward), and (ii) a single logical point of failure.
%
%
On the other hand, brokerless libraries 
omit the intermediary broker entirely. They provide enhanced socket abstractions that enable \emph{direct communication} among peers. Messages are byte sequences that the library queues internally between threads or processes and exposes at the destination socket.
As a result, latency is reduced to a single network hop and can approach the transport’s theoretical minimum, whereas persistence, retransmission, and higher‑level delivery guarantees must be implemented at the application layer, when required.

%% file: qualitative.tex
\section{Qualitative Analysis}\label{sec:qualitative_analysis}

In this section, we discuss five brokerless messaging libraries: ZeroMQ, NanoMsg, NNG, ZeroRPC, and YAMI4. We qualitatively assess them based on license, documentation, and community support. 

\begin{table*}[tb]
    \centering
    \caption{Qualitative characteristics of the considered brokerless messaging libraries.}
    \label{tab:qualitative}
    \begin{tabular}{lrrrrr}
        \toprule
        \textbf{Characteristic}        & \textbf{ZeroMQ} & \textbf{NanoMsg} & \textbf{NNG} & \textbf{ZeroRPC} & \textbf{YAMI4} \\ \midrule
        \textbf{Development language}  & C++             & C                & C            & Python           & C++, Objective C \\
        \textbf{Year of first release} & 2007            & 2012             & 2016         & 2012             & 2010           \\
        \textbf{Client bindings}       & C, C++, \ldots  & C, C++, \ldots   & C            & Python, node.js  & C++, \ldots    \\ \midrule
        \textbf{License}               & MPL‑2.0 \cite{libzmq_license} 
                                       & MIT \cite{nanomsg_license} 
                                       & MIT \cite{nng_license}
                                       & MIT \cite{zerorpc_license}
                                       & BSL or GPL \\
        \textbf{Documentation}         & Extensive \cite{zeromq_docs}      
                                       & Good \cite{nanomsg_docs}
                                       & Good \cite{d2018nng}         
                                       & Limited
                                       & Extensive \cite{yami4}       
                                       \\ \midrule
        \textbf{GitHub Stars  /  Forks  /  Issues}  
                                       & 10.1k  /  2.4k  /  1.5k 
                                       & 6.1k  /  1k  /  600  
                                       & 4k  /  500  /  1.1k  
                                       & 3.2k  /  300  /  100  
                                       & — 
                                       \\ 
        \textbf{GitHub Recent Commits  /  Issues} 
                                       & 0  /  20          & 0  /  2            & 100  /  18  & 2  /  0                & —              \\ 
        \bottomrule
    \end{tabular}
    
    \vspace{0.3em}
    \raggedright
    \footnotesize
    \textbf{Notes:} GitHub metrics were retrieved in April 2025. Commit and issue activity covers January--April 2025. Counts are rounded and intended as “at least”. GitHub activity considered for \textbf{ZeroMQ} refers specifically to \texttt{zeromq/libzmq}~\cite{libzmq_github}. For \textbf{NanoMsg}, it refers to \texttt{nanomsg/nanomsg}~\cite{nanomsg_github}. For \textbf{NNG}, it refers to \texttt{nanomsg/nng}~\cite{nng_github}. For \textbf{ZeroRPC}, it refers to \texttt{0rpc/zerorpc-python}~\cite{zerorpc_github}.
\end{table*}

Table~\ref{tab:qualitative} reports technical facts and practical aspects of the libraries considered.
For all, installation is straightforward, and is therefore omitted.
%
%
ZeroRPC offers little up-to-date documentation or active community support, and YAMI4 lacks a public code repository. By contrast, the activity counters show that NNG is the only library currently under active development.

In the following, we provide an overview of the candidates. 

\subsubsection{ZeroMQ} \cite{hintjens2013zeromq} is a C++ library known for its scalability and ease of use in building high-throughput applications. The project started in 2007 as an alternative to the Advanced Message Queuing Protocol (AMQP) and is currently one of the most popular brokerless messaging libraries. 
%
%
It supports data-centric communication through its PUB/SUB model, which includes advanced patterns for performance, reliability, state distribution, and monitoring. 
The library offers a unified socket API that abstracts over transport details, simplifying application development by decoupling messaging logic from the underlying communication layer. In particular, core operations --- bind, connect, send, and receive --- are transport-agnostic, with the user specifying the transport via the endpoint address.

\subsubsection{NanoMsg}~\cite{nanomsg_page} is a re-imagining of ZeroMQ, realized as a complete rewrite in C, focusing on a lower memory footprint and improved algorithm efficiency~\cite{nanomsg_rationale}.
NanoMsg keeps a simple, transport-agnostic socket API, while aiming to improve upon several design aspects of ZeroMQ --- particularly by targeting full POSIX compliance to facilitate development and maintenance.  
It is not in active development but only in sustaining mode. 

\subsubsection{NNG}
NNG~\cite{nng_page} builds upon the design principles of ZeroMQ and NanoMsg while addressing many of their limitations, as detailed in its rationale~\cite{nng_rationale}.  
It has a new implementation, focusing on scalability, portability, and robustness. 
Its API removes legacy constraints from POSIX socket interfaces, offering a cleaner and more portable abstraction and facilitating the development of new transports.  
NNG eliminates the complex, error-prone state machine architecture of NanoMsg and relaxes its file-descriptor-driven implementation, aiming to overcome the poor scalability of NanoMsg's threading model. To this end, NNG performs all I/O operations asynchronously, which also simplifies integration with non-blocking and event-driven applications.

\subsubsection{ZeroRPC}~\cite{zerorpc} extends the capabilities of ZeroMQ by providing integrated RPC support.
Developed in Python, ZeroRPC offers a dynamically typed service interface that simplifies the construction of service-oriented architectures. 
It leverages MessagePack for efficient serialization of data in a JSON-compatible format.
However, it only supports data-centric communication through a simplified PUB/SUB pattern. Additionally, the setup and usage might be more challenging due to the lack of clear documentation.
\subsubsection{YAMI4}~\cite{yami4} is a lightweight messaging framework, 
designed for real-time control and monitoring applications.
It emphasizes simplicity in message exchange and offers features suited for resource-constrained environments.
It provides a simple form of PUB/SUB via its value publisher concept, and it is limited to In-Process and TCP transports. 
Unlike other frameworks, YAMI4 distinguishes between physical connections and logical destinations, allowing different communication patterns to coexist over a single connection and avoiding early commitment to a specific pattern~\cite{yami4_vs_zeromq}.  
YAMI4's source code is available under a GPL license, with commercial licensing options upon request, but it \emph{does not have a public repository}, which may restrict broader community adoption.

We exclude YAMI4 and ZeroRPC from our quantitative analysis, as 
YAMI4 lacks support for Inter-Process communication and does not provide a public repository, whereas ZeroRPC is primarily focused on remote procedure calls.
Therefore, we proceed with a detailed analysis of ZeroMQ, NanoMsg, and NNG,
as they 
(i) provide a similar, simple, and efficient API,
(ii) are actively maintained, and 
(iii) with a large community support.

%% file: methodology.tex
\section{Methodology}\label{sec:methodology}

This section introduces the design of our benchmarking suite, developed to systematically compare the performance of messaging libraries.
We describe the benchmarking experiment and the architecture of our testing framework, designed to ensure a fair and unbiased comparison of library performance. Additionally, we outline the workload conditions and define the performance metrics considered in our evaluation.


We consider the communication transports described in Section~\ref{sec:bml}, namely \emph{In-Process}, \emph{Inter-Process}, and \emph{TCP}.
%
%
The focus is on the PUB/SUB communication, fixing a single, representative pattern --- the most common we have observed in practice for messaging frameworks. 
%
%
We consider the ZeroMQ, NanoMsg, and NNG libraries, as motivated in Section~\ref{sec:qualitative_analysis}. Library versions used are described in Table~\ref{tab:bml}.

\begin{table}[tb]
    \centering
    \caption{Versions and descriptions of the libraries.}
    \begin{tabular}{ll}
    \toprule
    \textbf{Library} & \textbf{Version and Description} \\
    \midrule
    \textbf{ZeroMQ} & libzmq (core API, v4.3.4),
                 \\ & CZMQ (High-level C Binding, v4.2.1) \\
    \textbf{NanoMsg} & libnanomsg (C core API, v1.1.5) \\
    \textbf{NNG} & libnng (C core API, v1.9.0) \\
    \bottomrule
    \end{tabular}
    \label{tab:bml}
\end{table}


\subsection{Parameters}\label{sec:parameters}

%
For our benchmark, we consider the following parameters:
\begin{itemize}
\item The \emph{number of subscribers} consuming the messages from the queue.
\item The \emph{number of messages sent} from the publisher to all the subscribers.
\item The \emph{publishing interval}, i.e., the amount of microseconds between each sending of a message.
\item The \emph{payload size}, i.e., the size of each message.
\item The \emph{publishing delay}, i.e., the time in milliseconds that the publisher waits before starting to publish messages.
\end{itemize}
These parameters are passed as arguments to the simulator, together with the transport and the library used for the communication.

\subsection{Figures of Merit}\label{sec:figures_of_merit}

We consider latency, throughput, jitter, and CPU and memory usage as the figures of merit for our comparison.

\begin{itemize}
    \item \emph{Latency} measures the delay of the system, crucial in real-time and streaming applications, and provides insight into the responsiveness of the communication framework.
    \item \emph{Throughput} represents the rate at which messages are successfully delivered, reflecting the efficiency of the system under various workload conditions.
    \item \emph{CPU and memory usage} reveals the resource cost associated with achieving the observed performance, a tradeoff especially relevant in systems with tight resource budgets or real-time constraints.
    \item \emph{Jitter} measures the variation in latency over time. This metric is important in real-time applications, as high jitter can lead to unpredictable behavior in systems that rely on timely message delivery.
\end{itemize}

For $M$ messages successfully received, we consider the overall distribution of \emph{one-way transmission latencies} $l_i$, with $i \in [1, M]$.
To compute the latency, we encode a UNIX timestamp in the payload of each message sent. If the message is received, we compute the latency as the difference between the receiver and the sender timestamps. Finally, upon completion of the simulation, we compute \emph{minimum}, \emph{average}, \emph{p90}, \emph{p99}, and \emph{maximum} latencies.

Secondly, we consider \emph{throughput} as the number of megabytes delivered per second. We compute it considering the number of messages received $M$, which all have the same fixed size $\mathcal{P}$, over the amount of seconds that passed from the first message sent to the last message received.

Then, we consider \emph{the average one-way packet delay variation}, introduced in the RFC 3393 IETF standard track document~\cite{rfc3393}, here defined as ``jitter'', and computed as the average of the absolute difference between consecutive transmission latencies:
\begin{equation}
    \text{jitter} = \frac{1}{M-1} \sum_{i=1}^{M-1} | l_i - l_{i-1} | .
\end{equation}


Finally, we consider \emph{median CPU} and \emph{median memory} usage percentages. 
To compute these metrics, we take discrete samples throughout the simulation and calculate their median values.
We leverage the \texttt{psutil} Python package~\cite{psutil} to sample these metrics at each time step.
For memory usage, we specifically measure the \emph{Unique Set Size (USS)}, which represents the amount of memory unique to a process and not shared with others~\cite{dalmasso2013survey}. 
%
To account for multiple threads or processes running simultaneously, we \emph{aggregate} CPU and memory usage across all processes at each time step.
Specifically, at each sampling time $t$, we sum the percentages for the CPU and USS memory usage over all $N$ processes.
Then, we report the median of the aggregated values as the final metric.

\subsection{Benchmarking Experiment}\label{sec:benchmarking_experiment}

In our experiments, we use one publisher and $\mathcal{S}$ subscribers.
The subscribers run first, whereas the publisher is executed after a delay of $\mathcal{D}$ milliseconds.
The publisher enqueues a total of $\mathcal{C}$ messages, each intended to be received by all subscribers. However, due to potential losses, $M \leq \mathcal{C}$ denotes the actual number of messages successfully received.
Each message carries a payload of $\mathcal{P}$ bytes. A timestamp for latency measurement is embedded in the message, and padded to the required size, thus introducing no extra overhead.
The publisher sends a message every $\mathcal{T}$ microseconds (publishing time interval). When $\mathcal{T}=0$, the publisher sends messages at the maximum achievable rate, without delay. Algorithm~\ref{pse:pub_sub_pseudocodes} summarizes the behavior of the publisher and subscribers.

\begin{algorithm*}[tb]
    \begin{minipage}{.495\textwidth}
    \begin{algorithmic}[1]
        \Function{Publisher}{$\mathcal{D}, \mathcal{C}, \mathcal{P}, \mathcal{T}$} 
            \State Sleep for $\mathcal{D}$ milliseconds
            \State Bind to endpoint

            \For{$i \gets 1 \text{ to } \mathcal{C}$}
                \State Insert current timestamp into payload
                \State Fill with `A's to reach size $\mathcal{S}$
                \State Publish payload to endpoint
                \State Sleep for $\mathcal{T}$ microseconds
            \EndFor
        \EndFunction
    \end{algorithmic}
    \end{minipage}
    \hfill
    \begin{minipage}{.495\textwidth}
    \begin{algorithmic}[1]
        \Function{Subscriber}{$\mathcal{C}$}
            \State Connect to endpoint
            \State Initialize timer and array to store latencies
            \For{$i \gets 1 \text{ to } \mathcal{C}$}
                \State Extract timestamp from message payload
                \State Compute latency using timestamp
                \State Store latency in the array
            \EndFor
            \State Process and report results
        \EndFunction
    \end{algorithmic}
    \end{minipage}
    \caption{Publisher (left) and subscriber (right) pseudocodes, for a single run of our benchmark.}
    \label{pse:pub_sub_pseudocodes}
\end{algorithm*}

\subsection{Benchmarking Suite}\label{sec:suite}

Our framework is designed to be general and extensible, offering a fair and comprehensive evaluation of library performance across different workload conditions.
\begin{figure}[tb]
    \centering
    \includegraphics[width=\linewidth]{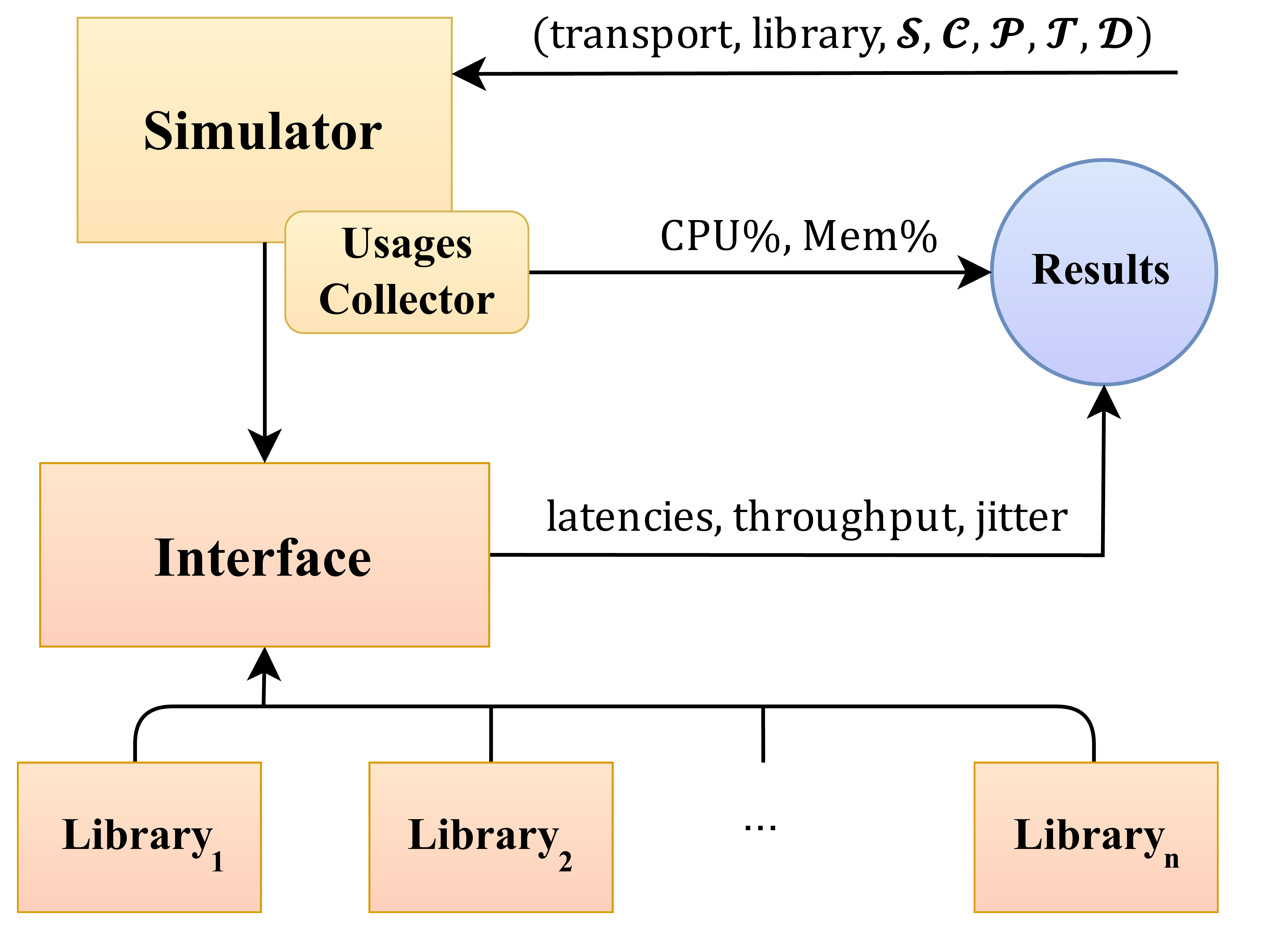}
    \caption{Overview of the benchmarking suite.}
    \label{fig:suite_architecture}
\end{figure}
Figure~\ref{fig:suite_architecture} shows a schematic overview of our framework components.

We conduct an extensive simulation, covering sets of the parameters detailed in Section~\ref{sec:parameters}.
All possible combinations of the parameter values are systematically explored, resulting in a total of $\prod_{i=1}^n |A_i|$ configurations, where $n$ is the number of parameters and $A_i$ is the set of values for the $i$-th parameter.

The \emph{simulator} module is responsible for running each experiment, corresponding to a combination of input parameters.
To provide reliable estimates of each library's performance, it applies two \emph{fixes}. 
First, it attaches the publisher and subscriber threads to isolated cores to avoid interference from other processes.
Second, it repeats the benchmarking experiment $R$ times and averages the results.
If multiple subscribers are involved, each producing different results, then the simulator first averages the results of the different subscribers, producing the result for a single run. Then, it averages over the results of multiple runs, producing the results for the given input parameters.
Furthermore, using the process identifiers of the threads running, the simulator implements the logic to sample the CPU and memory usage percentages at discrete time steps, as described in Section~\ref{sec:figures_of_merit}.

Finally, an \emph{interface} is built as an abstraction to the libraries' implementations of the basic operations (e.g., bind, connect, send, receive).
The interface implements the structure of the benchmarking experiment described in Section~\ref{sec:benchmarking_experiment}.
Any primitive operation is then implemented individually for each different library considered, inside a library-specific file.
As a result, to add a messaging library to our suite, it is sufficient to implement the basic functionalities required by the interface.
%
Algorithm~\ref{alg:suite} provides an abstract overview of our suite.

\begin{algorithm}[tb]
    \begin{algorithmic}[1]
        \Function{Benchmarking Suite}{$combs, R$}
            \For{$comb \in combs$}
                \State Initialize metrics cumulative variables
                \For{$r \gets 1 \text{ to } R$}
                    \State Run the exp. with parameters $comb$
                    \State Attach pub/sub threads to isolated cores
                    \State While running, extract (CPU/mem) usages
                    \State When done, extract results
                    \State Update cumulative variables
                \EndFor
                \State Process and report averages for $comb$
            \EndFor
        \EndFunction
    \end{algorithmic}
    \caption{Benchmarking suite pseudocode. 
    }
    \label{alg:suite}
\end{algorithm}

%% file: evaluation.tex
\section{Performance Comparison}\label{sec:performance_comparison}

In this section, we present the results of the performance analysis of ZeroMQ, NanoMsg, and NNG.\@ 

\subsection{Experimental Setup and Baseline Configuration}\label{sec:experimental_setup}\label{sec:baseline}

We run the benchmarking suite on a machine with the following specifications: Intel(R) Xeon(R) w3-2435 CPU with 8 cores at 3.1\,GHz,  and 64\,GB of RAM.
%


%
ZeroMQ, NanoMsg, and NNG are tested for the In-Process, Inter-Process, and TCP transports. 
In our benchmarks, we always run the publisher and subscribers on the same machine. 
%
The number of messages sent is fixed to $\mathcal{C}=5000$, and the starting delay of the publisher is set to $\mathcal{D}=1000$ milliseconds. 
We fix these parameters as they do not show significant variations in the performance of the libraries.
We focus on three key metrics: average latency, throughput, and CPU usage. To assess them, we conduct two main experiments --- one varying the message size, the other varying the number of subscribers.
When varying the size of messages, 
we fix the number of subscribers to $\mathcal{S}=1$ and vary the message size $P$ in power-of-two values, from 1\,KB to 512\,KB.
When varying the number of subscribers, we discuss scalability by fixing the message size to 32\,KB and changing the number of subscribers to 1, 2, 4, and 8.
For all the experiments, we used $R=4$ (see Section~\ref{sec:suite}).
\begin{table}[tb]
    \caption{Baseline configurations for the analysis. 
    }
    \begin{center}
    \begin{tabular}{lccc}
    \toprule
    \textbf{Parameter} & \multicolumn{3}{c}{\textbf{Configuration}} \\
    \cmidrule{2-4} 
    & \textbf{\textit{Latency}} & \textbf{\textit{\,\,\,\,\,\,\,\,\,Throughput}} & \textbf{\textit{CPU Usage}} \\
    \midrule
    \textbf{Publishing interval} ($\mathcal{T}$) & 1000\,µs & \,\,\,\,\,\,\,\,\,\,\,0\,µs & 0\,µs \\
    
    \textbf{Message size} ($\mathcal{P}$) & \multicolumn{3}{c}{32\,KB} \\
    
    \textbf{Subscribers} ($\mathcal{S}$) & \multicolumn{3}{c}{1}\\
    
    \textbf{Messages sent} ($\mathcal{C}$) & \multicolumn{3}{c}{5000} \\
    
    \textbf{Publisher delay} ($\mathcal{D}$) & \multicolumn{3}{c}{1000 ms} \\
    \bottomrule
    \end{tabular}
    \label{tab:configs}
    \end{center}
\end{table}

%
Given these conditions, we focus on the baseline settings for the publishing interval described in Table~\ref{tab:configs}.

We set the publishing interval to $T=0$ microseconds when measuring throughput, capturing the scenario in which the publisher sends messages at the maximum achievable rate (see Section~\ref{sec:benchmarking_experiment}). This avoids artificially limiting the throughput, as any non-zero value for $T$ would impose an upper bound on the rate of messages sent.
The same value of the publishing interval is used for the CPU usage analysis, as we want to consider the most resource-intensive case, where the publisher is sending messages at the fastest rate.
In contrast, for latency analysis, we deliberately use a publishing interval of $T=1000$ microseconds to decouple the latency measurements from queueing effects and inter-message interference. This ensures the observed latencies primarily reflect the library efficiency.

For all the results we report, each configuration is tested 
several times, and the results are averaged over the runs. This is done to guarantee a reliable estimate of the performance of the libraries, as detailed in Section~\ref{sec:suite}.

In Section~\ref{sec:results_other_metrics}, we briefly analyze the other metrics mentioned in Section~\ref{sec:figures_of_merit}. Full results are publicly available~\cite{suite}.

\subsection{In-Process Communication}\label{sec:results_inproc}

We analyze the performance of the libraries when the publisher and subscribers run in the same process.

\begin{figure*}[htb]
    \centering
    \begin{subfigure}[b]{0.325\linewidth}
        \includegraphics[width=\linewidth]{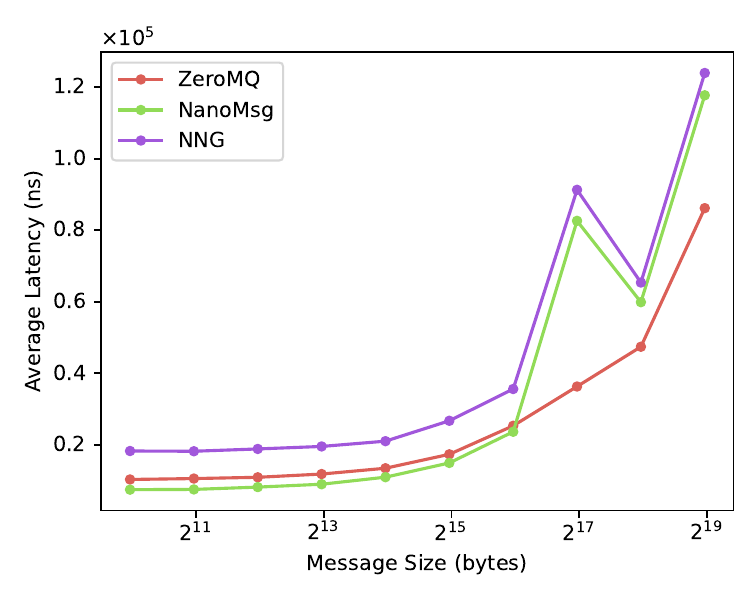}
        \caption{}\label{fig:inproc-latency-msgsize}
    \end{subfigure}
    \hfill
    \begin{subfigure}[b]{0.325\linewidth}
        \includegraphics[width=\linewidth]{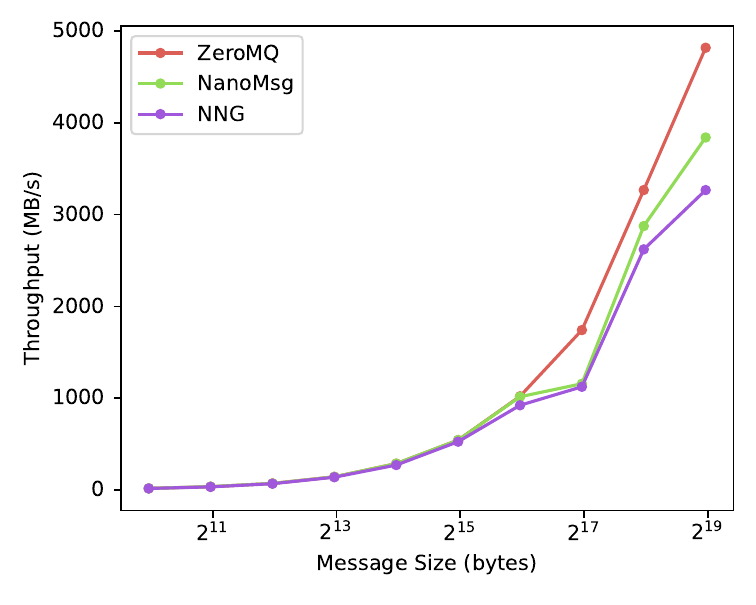}
        \caption{}\label{fig:inproc-throughput-msgsize}
    \end{subfigure}
    \hfill
    \begin{subfigure}[b]{0.325\linewidth}
        \includegraphics[width=\linewidth]{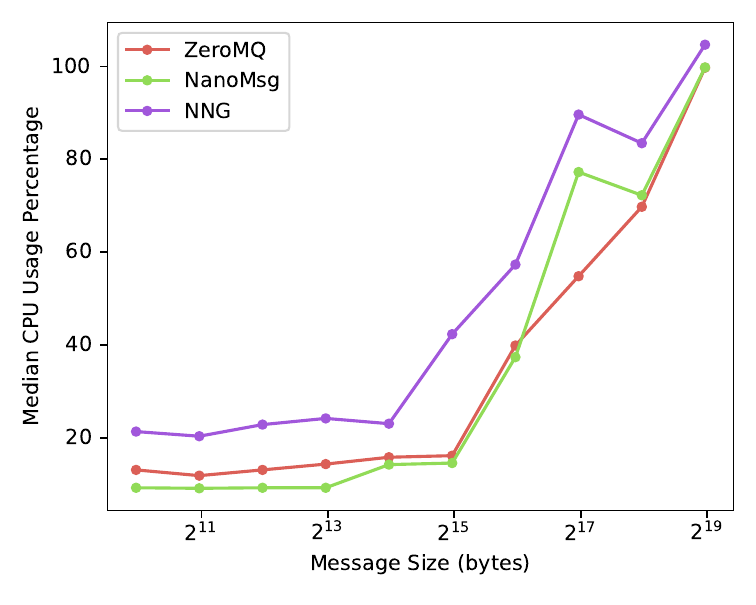}
        \caption{}\label{fig:inproc-cpu-msgsize}
    \end{subfigure}
    \caption{In-Process Communication: average latency (a), payload throughput (b), and CPU usage (c), when varying message size. The number of subscribers is fixed to 1.}
    \label{fig:inproc-msgsize}
\end{figure*}

\begin{figure*}[tb]
    \centering
    \begin{subfigure}[b]{0.325\linewidth}
        \includegraphics[width=\linewidth]{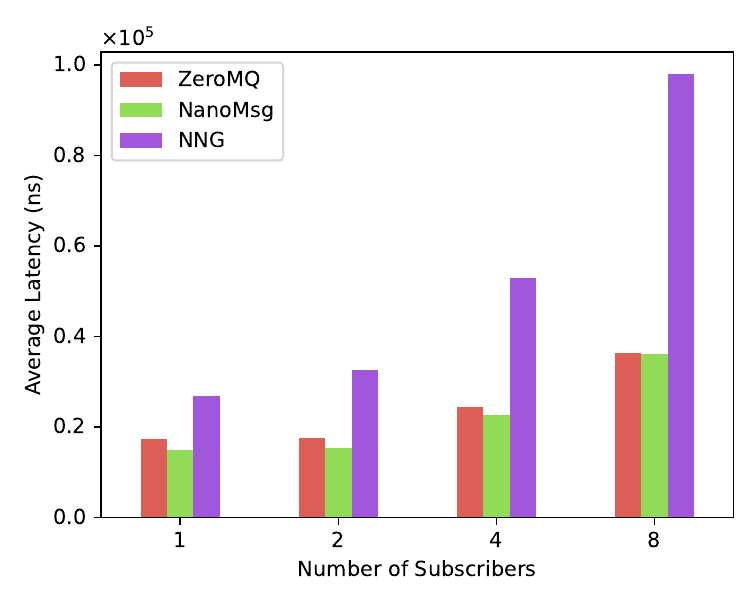}
        \caption{}\label{fig:inproc-latency-nsubs}
    \end{subfigure}
    \hfill
    \begin{subfigure}[b]{0.325\linewidth}
        \includegraphics[width=\linewidth]{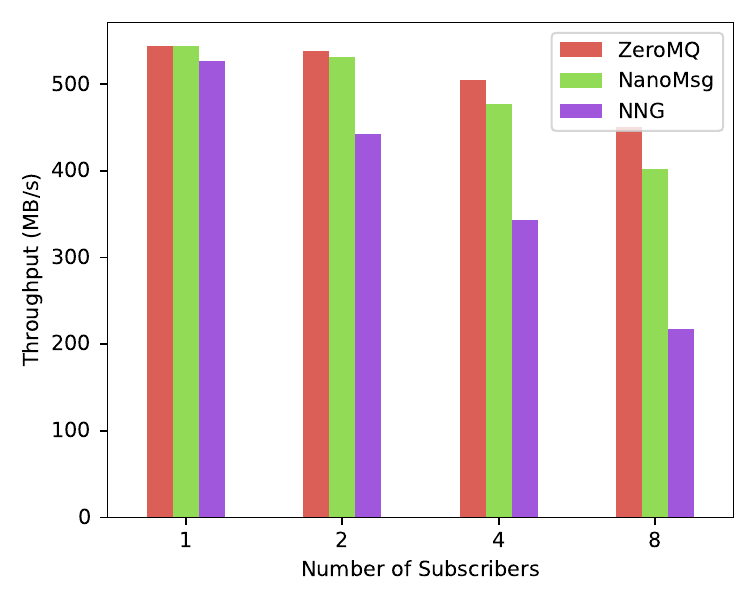}
        \caption{}\label{fig:inproc-throughput-nsubs}
    \end{subfigure}
    \hfill
    \begin{subfigure}[b]{0.325\linewidth}
        \includegraphics[width=\linewidth]{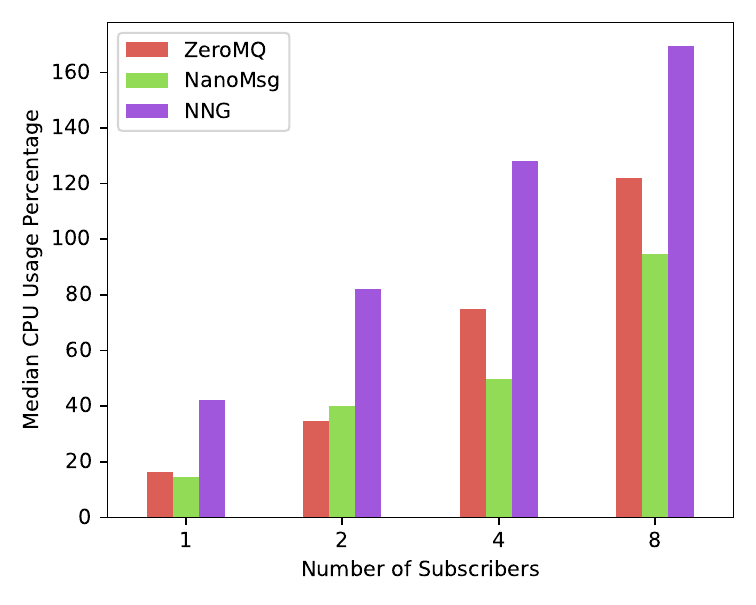}
        \caption{}\label{fig:inproc-cpu-nsubs}
    \end{subfigure}
    \caption{In-Process Communication: average latency (a), payload throughput (b), and CPU usage (c), when varying the number of subscribers. The message size is fixed to 32\,KB.}
    \label{fig:inproc-nsubs}
\end{figure*}

Figure~\ref{fig:inproc-msgsize} shows the results when fixing the number of subscribers $\mathcal{S}=1$ and varying the message size, according to the baseline values described in Section~\ref{sec:baseline}.
%
The measured latencies (Figure~\ref{fig:inproc-latency-msgsize}) range from \emph{less than 10 microseconds} to over one hundred microseconds, mainly depending on the message size.
As the message size increases, the average latency also increases, as expected due to the additional processing time required to handle larger messages. In particular, at 128\,KB, NanoMsg and NNG exhibit a peak in latency, after which their performance degrades.

\begin{observation}\label{obs:inproc-latency-payload}
\textit{NanoMsg has the lowest latency with small payloads, while ZeroMQ has the lowest latency once payloads become larger.}
\end{observation}


Figures~\ref{fig:inproc-throughput-msgsize} and \ref{fig:inproc-cpu-msgsize}
present the throughput and CPU usage results.
The throughput increases as the size of the payload increases, reaching a peak at 512\,KB. This is expected, as larger payloads can be sent in fewer messages, reducing the overhead associated with message passing. The highest throughput achieved by the libraries ranges \emph{from 3 to 5 gigabytes per second}. 
In tradeoff with the throughput, we notice a higher CPU consumption as the payload becomes larger, as expected due to the increased amount of bytes to process per time unit.
ZeroMQ achieves the highest throughput ($\approx 4.8\,\mathrm{GB/s}$). 
NanoMsg and NNG monotonically increase but peak at lower values, respectively at $\approx 3.8\,\mathrm{GB/s}$ and $\approx 3.3\,\mathrm{GB/s}$.
Regarding the cumulative median CPU usage (see Section~\ref{sec:figures_of_merit}), we observe that NanoMsg has the lowest CPU consumption for smaller payloads --- up to 128\,KB --- while still maintaining a competitive throughput. However, as the message size increases, ZeroMQ becomes the most efficient library, achieving the highest throughput with the lowest CPU consumption. NNG shows lower throughput with higher latency and CPU usage than the other two libraries.

\begin{observation}\label{obs:inproc-payload}
\emph{NanoMsg shows the best latency, throughput, and CPU results up to 64\,KB, after which ZeroMQ overtakes}.
\end{observation}

Figure~\ref{fig:inproc-nsubs}, shows the results for the same metrics when the number of subscribers increases. As mentioned, the message size is fixed to 32\,KB for these experiments.
Figure~\ref{fig:inproc-latency-nsubs}, shows that latency grows with the number of subscribers, especially for NNG. At the same time, ZeroMQ closes the gap with NanoMsg.
We observe similar insights for throughput (Figure~\ref{fig:inproc-throughput-nsubs}), which decreases as the subscribers increase. NNG is competitive for one subscriber, however it does not scale as well as the other libraries.
ZeroMQ is the library that scales best, outperforming NanoMsg.
The CPU usage result (Figure~\ref{fig:inproc-cpu-nsubs}) highlights that the usage percentage grows as the number of subscribers increases, as expected. The cumulative percentage (see Section~\ref{sec:figures_of_merit}) reaches peaks above 100\% for ZeroMQ and NNG, whereas NanoMsg's CPU usage scales the best for $\mathcal{S}=\{4,8\}$.

\begin{observation}\label{obs:inproc-subs}
\textit{ZeroMQ scales best in terms of throughput, whereas NanoMsg scales best  in terms of CPU usage.}
\end{observation}

\subsection{Inter-Process Communication}\label{sec:results_ipc}
\begin{figure*}[htb]
    \centering
    \begin{subfigure}[b]{0.325\linewidth}
        \centering
        \includegraphics[width=\linewidth]{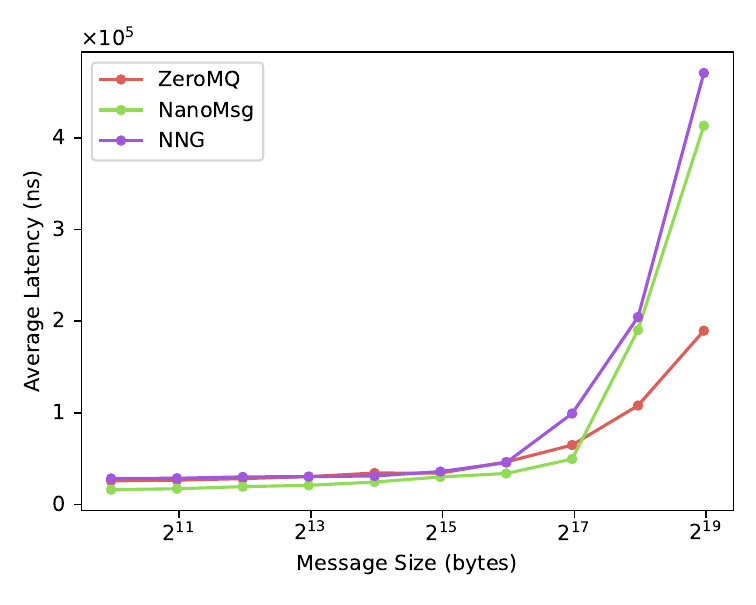}
        \caption{}\label{fig:ipc-latency-msgsize}
    \end{subfigure}
    \hfill
    \begin{subfigure}[b]{0.325\linewidth}
        \centering
        \includegraphics[width=\linewidth]{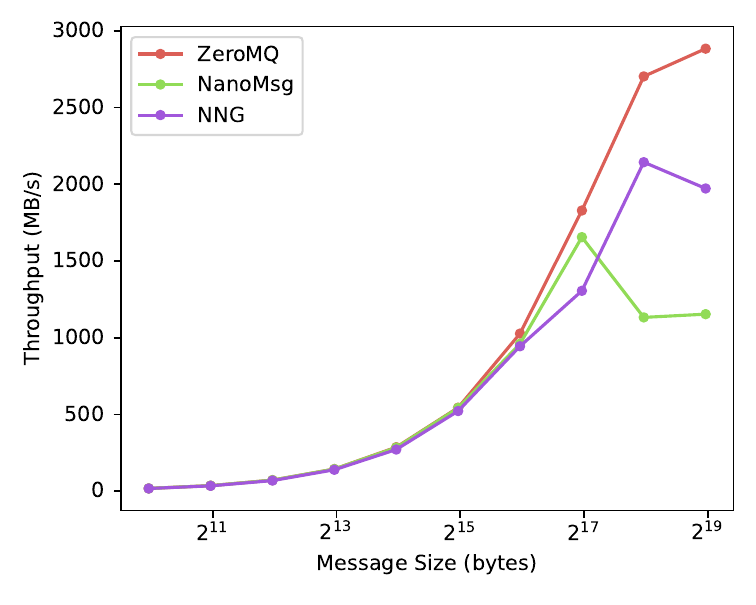}
        \caption{}\label{fig:ipc-throughput-msgsize}
    \end{subfigure}
    \hfill
    \begin{subfigure}[b]{0.325\linewidth}
        \centering
        \includegraphics[width=\linewidth]{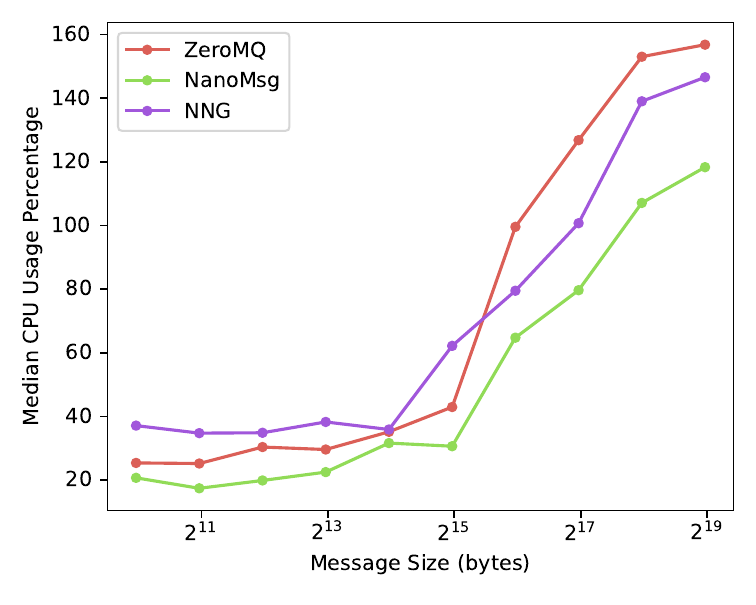}
        \caption{}\label{fig:ipc-cpu-msgsize}
    \end{subfigure}
    
    \caption{Inter-Process Communication: average latency (a), payload throughput (b), and CPU usage (c), when varying message size. The number of subscribers is fixed to 1.}
    \label{fig:ipc-msgsize}
\end{figure*}

\begin{figure*}[tb]
    \centering
    \begin{subfigure}[b]{0.325\linewidth}
        \centering
        \includegraphics[width=\linewidth]{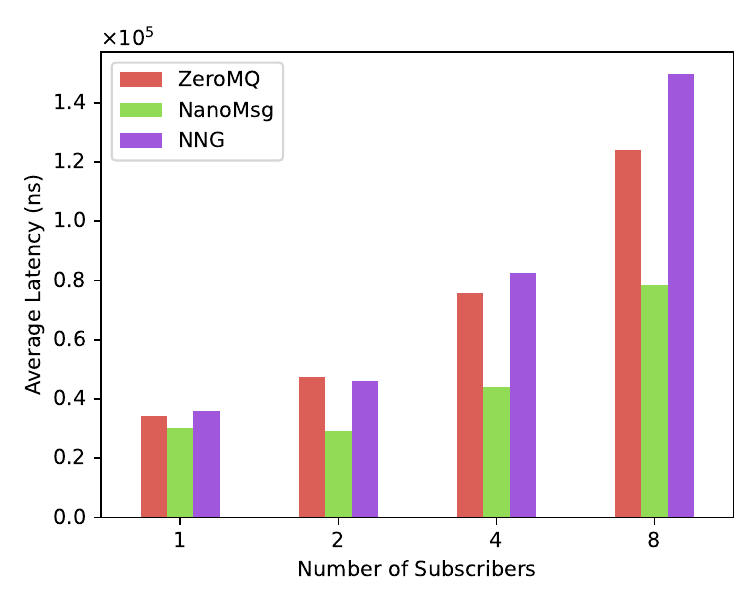}
        \caption{}\label{fig:ipc-latency-nsubs}
    \end{subfigure}
    \hfill
    \begin{subfigure}[b]{0.325\linewidth}
        \centering
        \includegraphics[width=\linewidth]{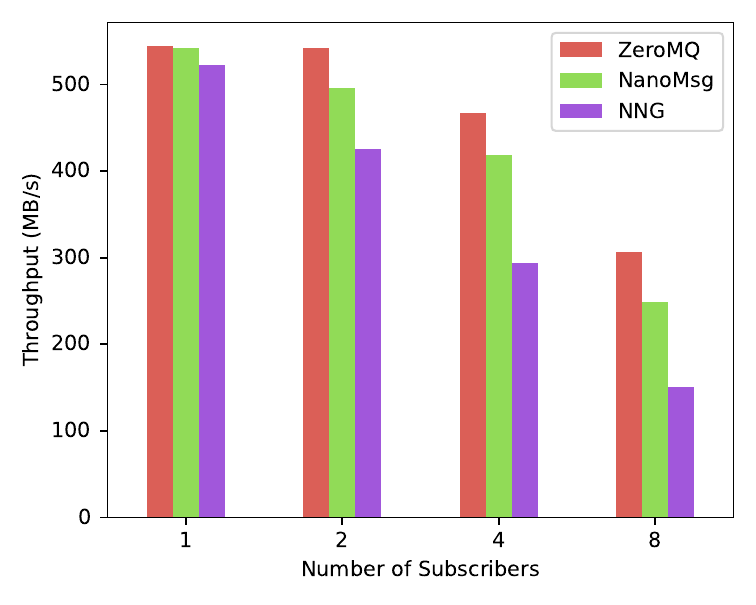}
        \caption{}\label{fig:ipc-throughput-nsubs}
    \end{subfigure}
    \hfill
    \begin{subfigure}[b]{0.325\linewidth}
        \centering
        \includegraphics[width=\linewidth]{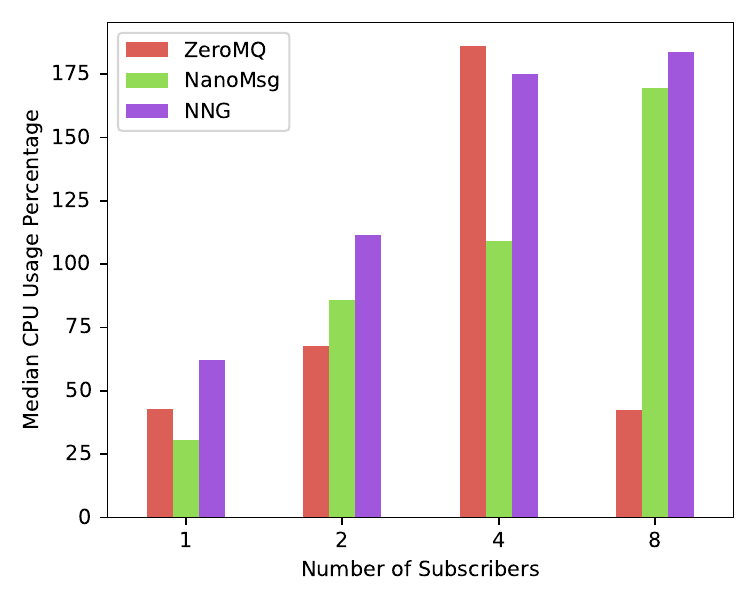}
        \caption{}\label{fig:ipc-cpu-nsubs}
    \end{subfigure}
    
    \caption{Inter-Process Communication: average latency (a), payload throughput (b), and CPU usage (c), when varying the number of subscribers. The message size is fixed to 32\,KB.}
    \label{fig:ipc-nsubs}
\end{figure*}

We repeat the same experiments considering the Inter-Process transport, showing the results in Figure~\ref{fig:ipc-msgsize} and Figure~\ref{fig:ipc-nsubs}.

Figure~\ref{fig:ipc-msgsize} depicts the results obtained when varying the message size.
Average latencies (shown in Figure~\ref{fig:ipc-latency-msgsize}) 
range from \emph{under 20\,µs} at 1 KB to over 400\,µs at 512 KB, corresponding to 2.5x-4x the In-Process values.

\begin{observation}\label{obs:ipc-latency-payload}
\textit{
For Inter-Process communication, NanoMsg achieves the best latency for a message size of up to 128\,KB, after which ZeroMQ performs considerably better.}
\end{observation}

Throughput results are shown in Figure~\ref{fig:ipc-throughput-msgsize}. Similar to the In‑Process results, the libraries reach throughput peaks \emph{from 1 to 3 gigabytes per second}, with ZeroMQ achieving the highest value ($\approx 2.9\,\mathrm{GB/s}$).
NNG shows a higher throughput than NanoMsg for the two largest message sizes.
The CPU usage results are shown in Figure~\ref{fig:ipc-cpu-msgsize}, indicating that cumulative utilization percentages are up to 1.6 times higher than those observed in the In-Process scenario.
ZeroMQ uses less CPU than NNG up to 32\,KB, whereas NNG becomes more CPU-efficient for larger payloads.

\begin{observation}\label{obs:ipc-cpu-payload}
\textit{
For Inter-Process communication, NanoMsg consistently shows the lowest CPU usage for all message sizes.}
\end{observation}


Figure~\ref{fig:ipc-latency-nsubs} shows that, having fixed the message size to 32\,KB, as the number of subscribers increases, latency tends to grow.
This is more evident for ZeroMQ and NNG, while NanoMsg demonstrates significantly better scalability. 
In terms of throughput (Figure~\ref{fig:ipc-throughput-nsubs}), the three libraries show similar performance when only one subscriber is present. However, as the number of subscribers increases, ZeroMQ achieves higher throughput compared to NNG and NanoMsg, with NNG exhibiting a more pronounced decline.
Regarding the CPU usage (Figure~\ref{fig:ipc-cpu-nsubs}), the results exhibit non‑linear behavior, with each configuration revealing distinct patterns. NanoMsg performs best in scenarios with 1 and 4 subscribers, while ZeroMQ outperforms the others in configurations with 2 and 8 subscribers.

\begin{observation}\label{obs:ipc-subs}
\textit{
For Inter-process communication, NanoMsg scales best for average latency, whereas ZeroMQ scales best for throughput. ZeroMQ and NanoMsg, depending on the number of subscribers, are the most CPU efficient libraries.}
\end{observation}

\subsection{TCP Communication}\label{sec:results_tcp}
This section evaluates the performance of libraries considering
TCP-based communication.
We replicate the same experiments conducted for In- and Inter-Process communication, configuring the libraries to use the TCP transport.
This results in additional overhead due to the traversal of the network stack, even though the messages are routed through the loopback interface.
We restrict the TCP experiments to a single host so that (i) micro-second–level one-way latencies can be measured without requiring sub-$\mu s$ clock synchronization between distinct machines, and (ii) any variability introduced by network interface cards, cables, or switches is eliminated, allowing us to isolate the overhead of traversing the kernel’s TCP/IP stack.

\begin{figure*}[htb]
    \centering
    \begin{subfigure}[b]{0.325\linewidth}
        \centering
        \includegraphics[width=\linewidth]{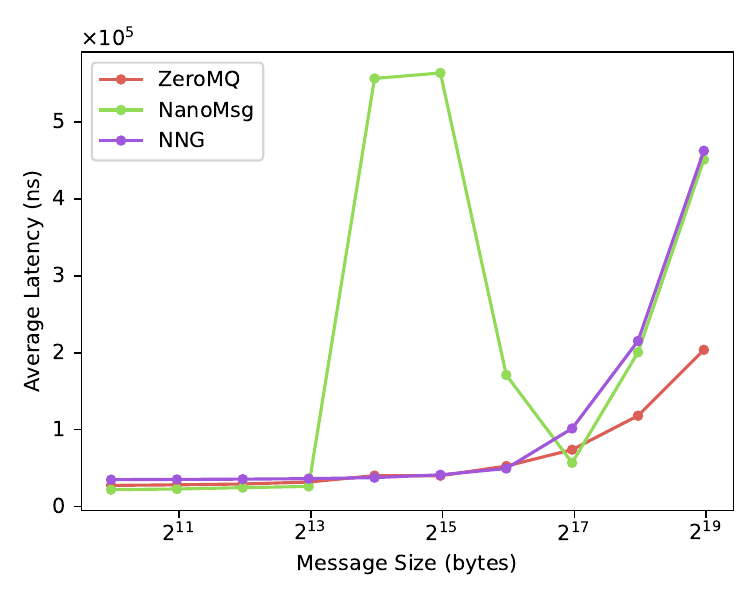}
        \caption{}\label{fig:tcp-latency-msgsize}
    \end{subfigure}
    \hfill
    \begin{subfigure}[b]{0.325\linewidth}
        \centering
        \includegraphics[width=\linewidth]{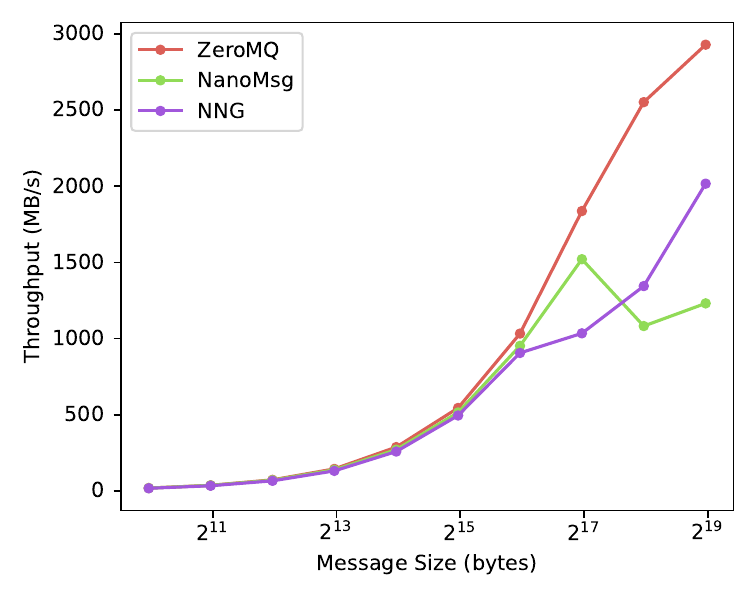}
        \caption{}\label{fig:tcp-throughput-msgsize}
    \end{subfigure}
    \hfill
    \begin{subfigure}[b]{0.325\linewidth}
        \centering
        \includegraphics[width=\linewidth]{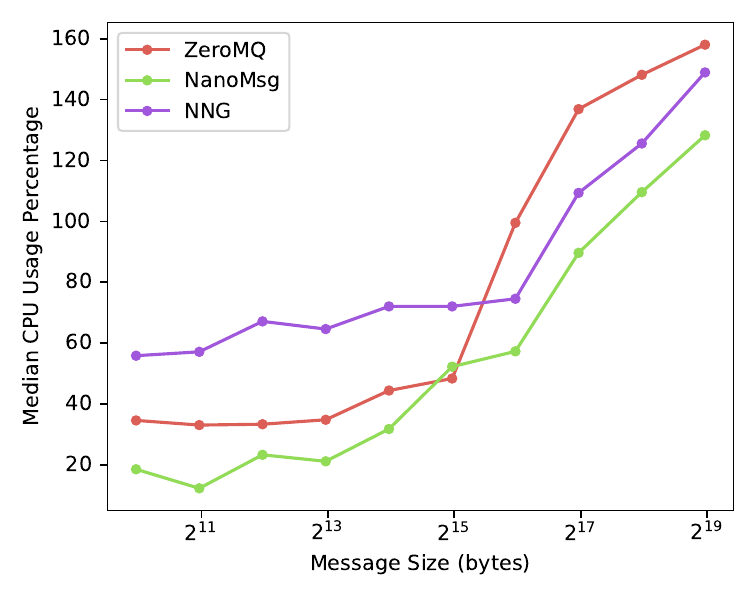}
        \caption{}\label{fig:tcp-cpu-msgsize}
    \end{subfigure}

    \caption{TCP Communication: average latency (a), payload throughput (b), and CPU usage (c), when varying the message size. The number of subscribers is fixed to 1.}
    \label{fig:tcp-msgsize}
\end{figure*}

\begin{figure*}[tb]
    \centering
    \begin{subfigure}[b]{0.325\linewidth}
        \centering
        \includegraphics[width=\linewidth]{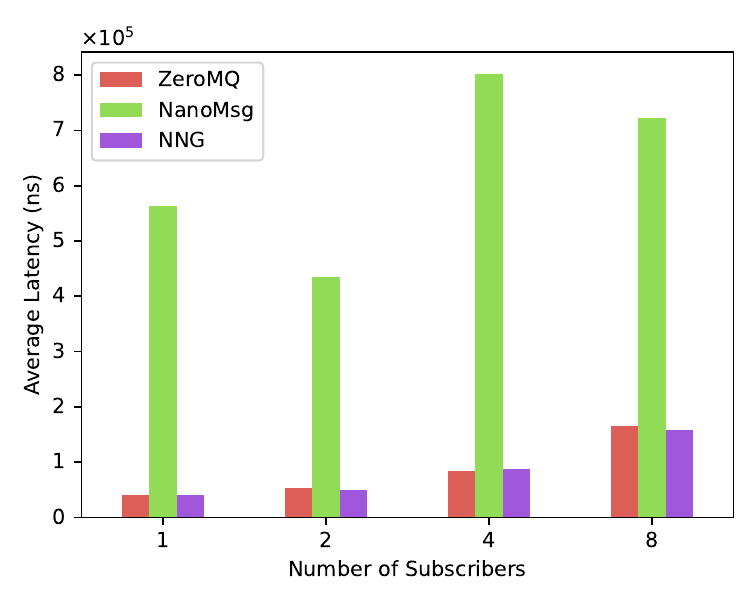}
        \caption{}\label{fig:tcp-latency-nsubs}
    \end{subfigure}
    \hfill
    \begin{subfigure}[b]{0.325\linewidth}
        \centering
        \includegraphics[width=\linewidth]{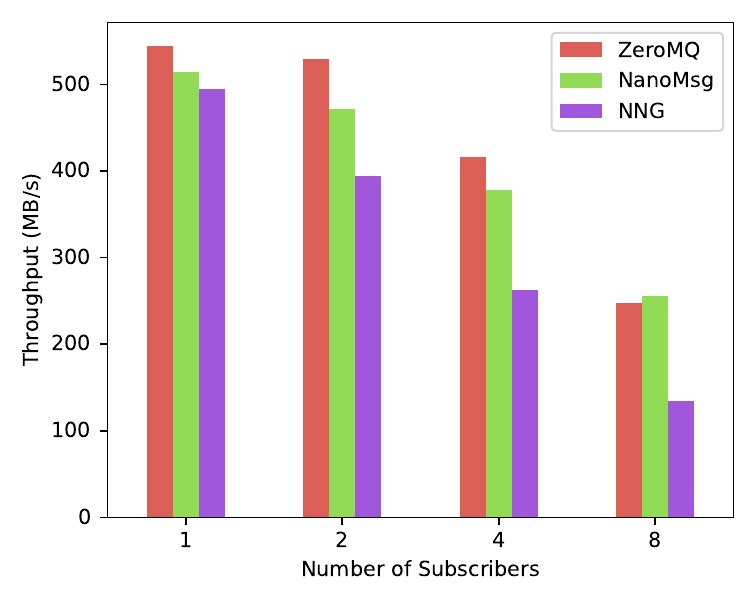}
        \caption{}\label{fig:tcp-throughput-nsubs}
    \end{subfigure}
    \hfill
    \begin{subfigure}[b]{0.325\linewidth}
        \centering
        \includegraphics[width=\linewidth]{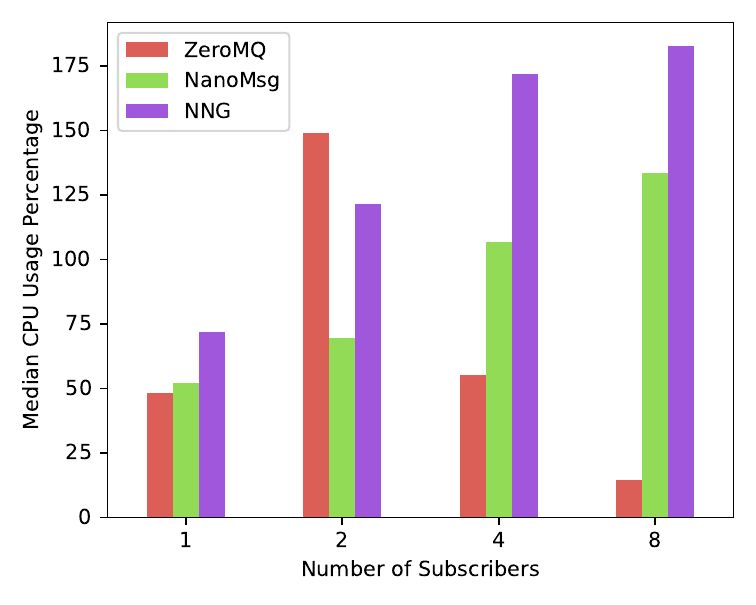}
        \caption{}\label{fig:tcp-cpu-nsubs}
    \end{subfigure}

    \caption{TCP Communication: average latency (a), payload throughput (b), and CPU usage (c), when varying the number of subscribers. The message size is fixed to 32\,KB.}
    \label{fig:tcp-nsubs}
\end{figure*}

Analyzing average latency as a function of payload size (Figure~\ref{fig:tcp-latency-msgsize}), we observe latency values in a range \emph{comparable to those of Inter‑Process communication}. This suggests that the overhead of the TCP stack is relatively low. 
NanoMsg has the lowest latency up to 8\,KB, and then exhibits a strange behavior, up to 64\,KB, possibly due to an internal limitation of the library. 
%
%
In this region, NNG and ZeroMQ achieve the lowest latency. ZeroMQ becomes the best choice for large message sizes (greater than 128\,KB).

Throughput performance (Figure~\ref{fig:tcp-throughput-msgsize}) shows similar results to the one presented for the Inter‑Process scenario. 

\begin{observation}\label{obs:tcp-thr-payload}
\emph{For TCP communication, ZeroMQ achieves the highest peak throughput ($\approx 2.9\,\mathrm{GB/s}$), outperforming NanoMsg and NNG for message sizes higher than 128\,KB.}
\end{observation}

NanoMsg throughput achieves a peak around 1.5\,GB/s for 128\,KB but then decreases, while NNG throughput monotonically increases, surpassing NanoMsg for large payloads, and achieving a peak around 2\,GB/s. 
The CPU usage analysis (Figure~\ref{fig:tcp-cpu-msgsize}) indicates that \emph{NanoMsg generally offers the lowest CPU usage}, with an exception at 32\,KB where ZeroMQ marginally outperforms it. ZeroMQ and NNG follow, for message sizes respectively lower and higher than 32\,KB.


In Figure~\ref{fig:tcp-nsubs}, we examine the impact of scaling the number of subscribers at a fixed payload size of 32\,KB.
The latency behavior (Figure~\ref{fig:tcp-latency-nsubs}) shows that the peak in latency exhibited by NanoMsg is also present for a higher number of subscribers. On the other hand, \emph{ZeroMQ and NNG offer relatively scalable latencies} across varying subscriber counts.
Throughput scalability (Figure~\ref{fig:tcp-throughput-nsubs}) presents a similar trend with respect to the other transports considered. ZeroMQ maintains the highest throughput as the subscriber count increases, except for the scenario with 8 subscribers, where NanoMsg surpasses both ZeroMQ and NNG, providing a higher throughput. 
Finally, we consider the CPU usage as the subscriber count varies (Figure~\ref{fig:tcp-cpu-nsubs}). 
As observed in the Inter‑Process case, ZeroMQ exhibits non‑linear trends. 
It emerges as the most CPU‑efficient library with 1, 4, and 8 subscribers, whereas NanoMsg is more efficient with 2 subscribers. 

\begin{observation}\label{obs:tcp-subs}
\emph{For TCP communication, as subscriber count grows, ZeroMQ and NNG scale best for latency, NanoMsg for throughput; the most CPU-efficient library is, depending on the scenario, either ZeroMQ or NanoMsg.}
\end{observation}

\subsection{Other Metrics}\label{sec:results_other_metrics}
The goal of this section is to discuss some other interesting results that we observed during the evaluation.
We focus on minimum, maximum, and percentile values for the latency, together with jitter and memory usage.
Due to space constraints, we only present the results for the Inter-Process transport; the complete results for every other transport are available in~\cite{suite}.

\begin{figure*}[tb]
    \centering
    \begin{subfigure}[b]{0.325\linewidth}
        \centering
        \includegraphics[width=\linewidth]{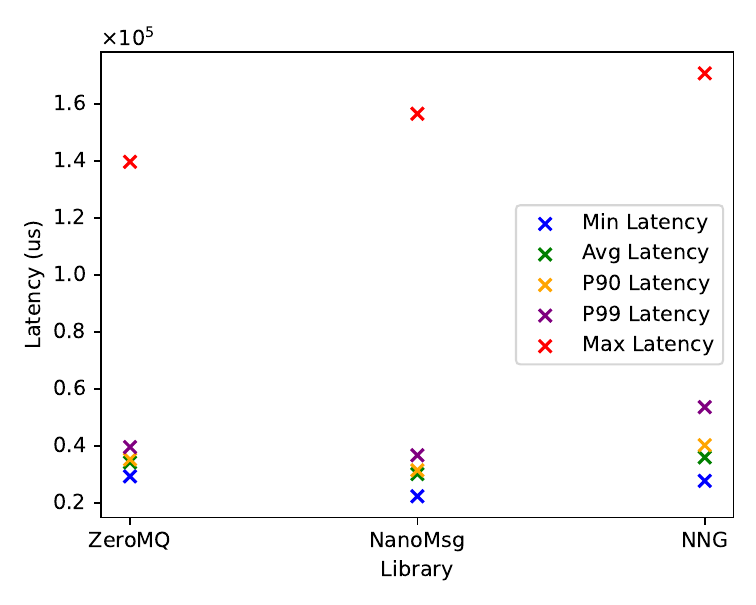}
        \caption{Distribution of latencies for each library, considering one subscriber, message size fixed to 32\,KB, and publishing interval of 1000µs. 
        } 
        \label{fig:ipc-latency-dist}
    \end{subfigure}
    \hfill
    \begin{subfigure}[b]{0.325\linewidth}
        \centering
        \includegraphics[width=\linewidth]{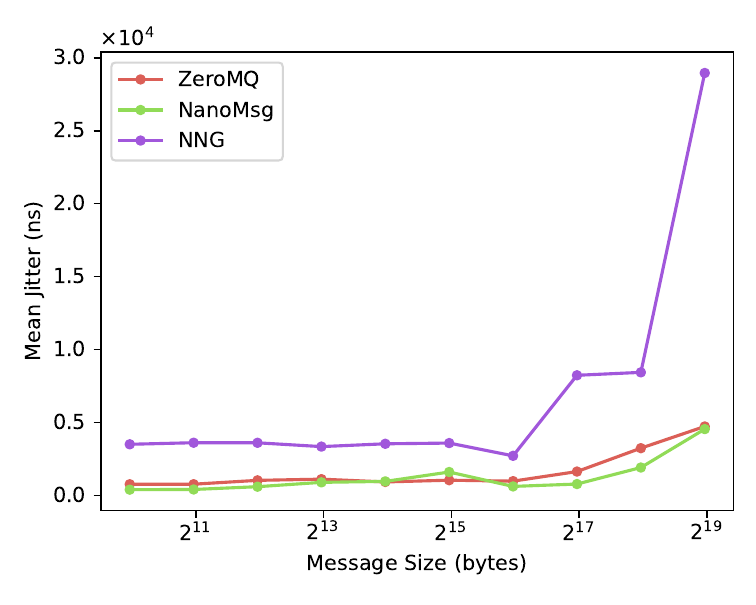}
        \caption{Jitter. We consider one subscriber and the publishing interval fixed to 1000\,µs, varying the message size. 
        }
        \label{fig:ipc-jitter}
    \end{subfigure}
    \hfill
    \begin{subfigure}[b]{0.325\linewidth}
        \centering
        \includegraphics[width=\linewidth]{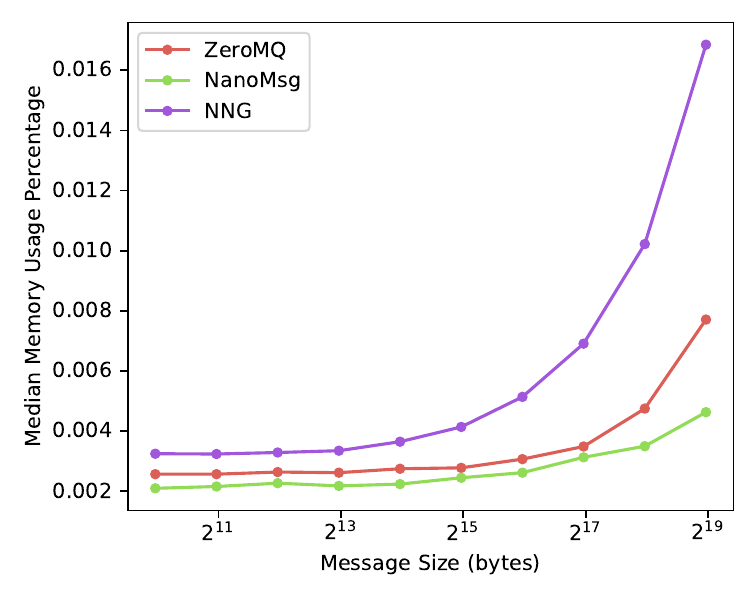}
        \caption{Memory usage. We consider one subscriber and the publishing interval fixed to 1000\,µs, varying the message size. 
        }
        \label{fig:ipc-memory}
    \end{subfigure}
    \caption{Inter-Process Communication: results for the jitter, memory usage, and latency values at different percentiles. }
\end{figure*}

First, we further discuss latencies by showing more details about their overall distribution over a single configuration.
In particular, from the baseline setup detailed in Section~\ref{sec:baseline}, we fix the value of the message size to 32\,KB, thus considering a single set of values for the input parameters. 

For this setting, Figure~\ref{fig:ipc-latency-dist} shows the minimum, average, p90, p99, and maximum latencies for each library.
The plot highlights that, while most latency values (minimum, average, p90, and p99) are close, \emph{the maximum latency is significantly higher}.
ZeroMQ exhibits the narrowest latency distribution, achieving the best result in terms of maximum latency.
NanoMsg achieves the best average latency and also performs best for the minimum, p90, and p99 metrics.
In contrast, NNG shows the widest latency distribution and exhibits the worst results for the maximum latency.

\begin{observation}\label{obs:ipc-lat-dist}
\emph{ZeroMQ minimizes worst‑case latency, whereas NanoMsg is best on average results.} 
\end{observation}

Second, we analyze jitter, 
using the configuration with a publishing interval of 1000\,µs as presented in Section~\ref{sec:baseline}. The results are shown in Figure~\ref{fig:ipc-jitter}.
NanoMsg and ZeroMQ exhibit the best performance in terms of jitter, maintaining average variations between consecutive message latencies \emph{in the order of one microsecond}.
In contrast, NNG shows a higher jitter for all message sizes, still remaining below ten microseconds on average, except for a peak at the largest payload.

\begin{observation}\label{obs:ipc-jitter}
\emph{NanoMsg and ZeroMQ keep the jitter near 1\,µs, whereas NNG shows higher values and spikes for higher payloads, e.g., 512\,KB.}
\end{observation}

Finally, in Figure~\ref{fig:ipc-jitter}, we consider memory usage.
We can see that the memory usage for all libraries increases monotonically as the message size increases, as expected. However, all the libraries show \emph{very low memory consumption}, under $0.01\%$ in most cases.
In particular, NanoMsg achieves the lowest consumption, followed by ZeroMQ and NNG.

\begin{observation}\label{obs:ipc-mem}
\emph{The memory usage for all three libraries is relatively low, with NanoMsg being the most memory efficient.}
\end{observation}

\subsection{Optimality Regions}\label{sec:results_optimality}

Finally, we explore every combination of the discrete sets of subscriber number and message size considered so far and identify, for each region of the parameter space, the library that is \emph{optimal}, i.e., performs best according to our figures of merit, \emph{within the scope of the three frameworks evaluated}. 

\begin{figure*}[tb]
    \centering
    \begin{subfigure}[b]{0.30\linewidth}
        \centering
        \includegraphics[width=\linewidth]{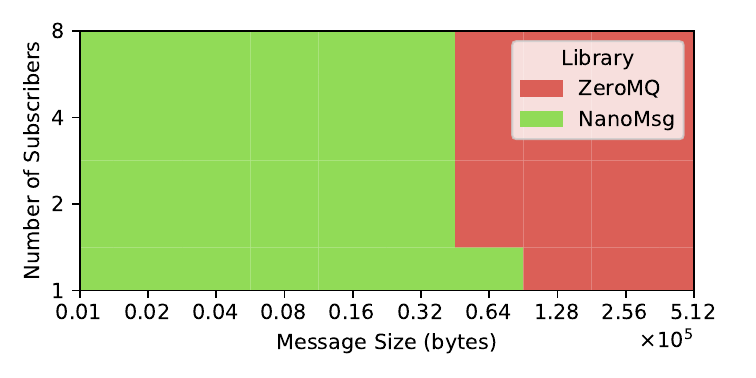}
        \caption{}\label{fig:opt-inproc-latency}
    \end{subfigure}
    \hfill
    \begin{subfigure}[b]{0.30\linewidth}
        \centering
        \includegraphics[width=\linewidth]{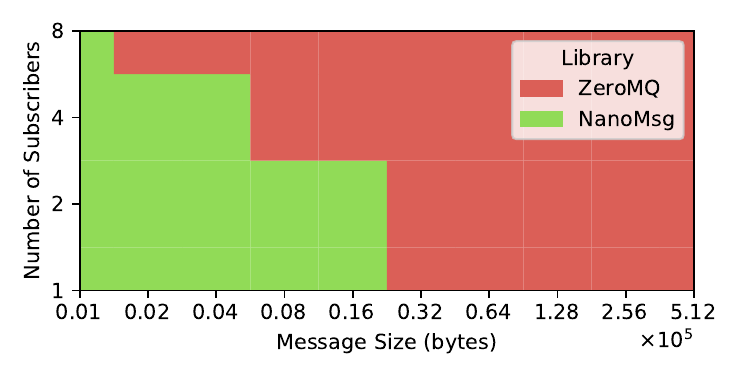}
        \caption{}\label{fig:opt-inproc-throughput}
    \end{subfigure}
    \hfill
    \begin{subfigure}[b]{0.30\linewidth}
        \centering
        \includegraphics[width=\linewidth]{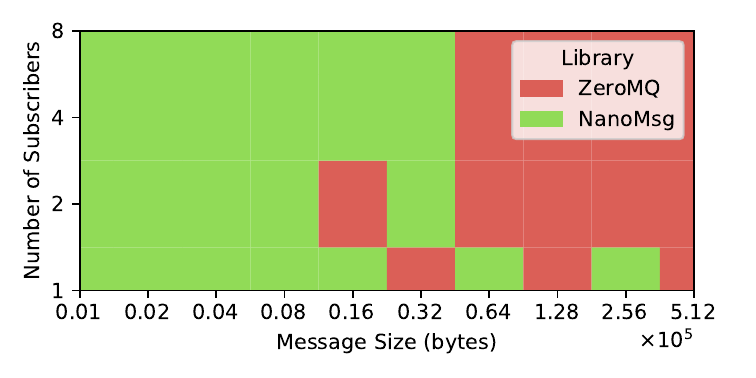}
        \caption{}\label{fig:opt-inproc-cpu}
    \end{subfigure}

    \begin{subfigure}[b]{0.30\linewidth}
        \centering
        \includegraphics[width=\linewidth]{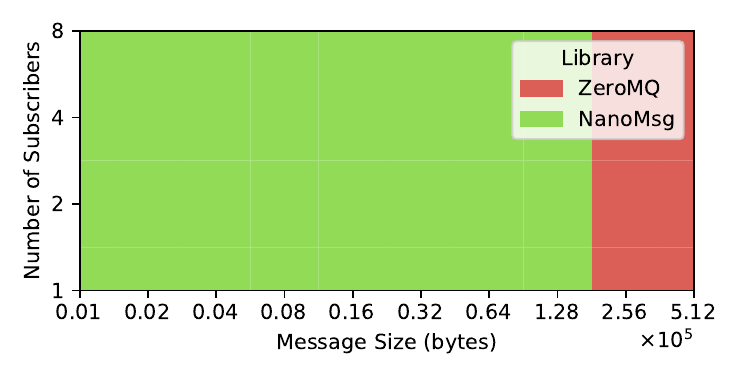}
        \caption{}\label{fig:opt-ipc-latency}
    \end{subfigure}
    \hfill
    \begin{subfigure}[b]{0.30\linewidth}
        \centering
        \includegraphics[width=\linewidth]{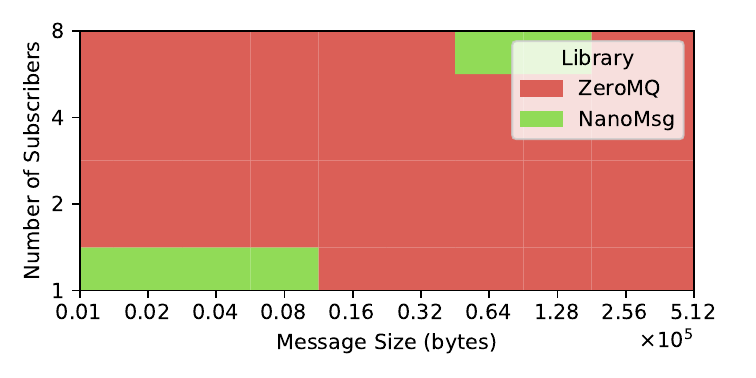}
        \caption{}\label{fig:opt-ipc-throughput}
    \end{subfigure}
    \hfill
    \begin{subfigure}[b]{0.30\linewidth}
        \centering
        \includegraphics[width=\linewidth]{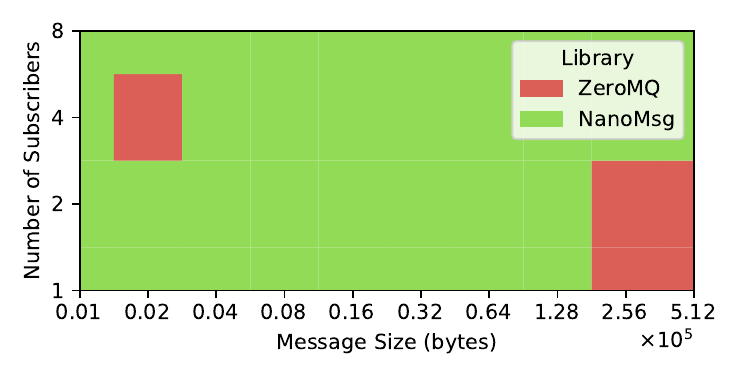}
        \caption{}\label{fig:opt-ipc-cpu}
    \end{subfigure}

    \begin{subfigure}[b]{0.30\linewidth}
        \centering
        \includegraphics[width=\linewidth]{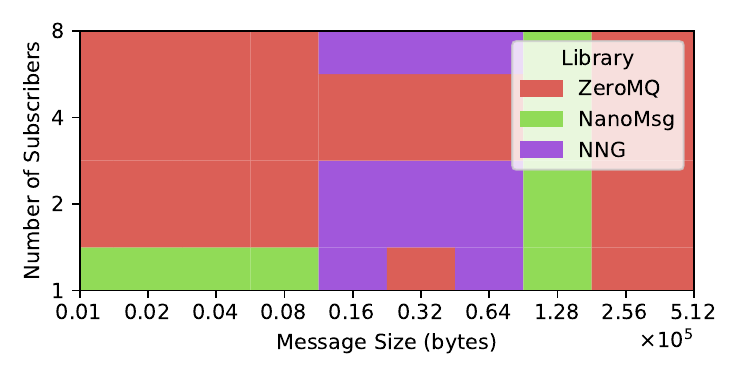}
        \caption{}\label{fig:opt-tcp-latency}
    \end{subfigure}
    \hfill
    \begin{subfigure}[b]{0.30\linewidth}
        \centering
        \includegraphics[width=\linewidth]{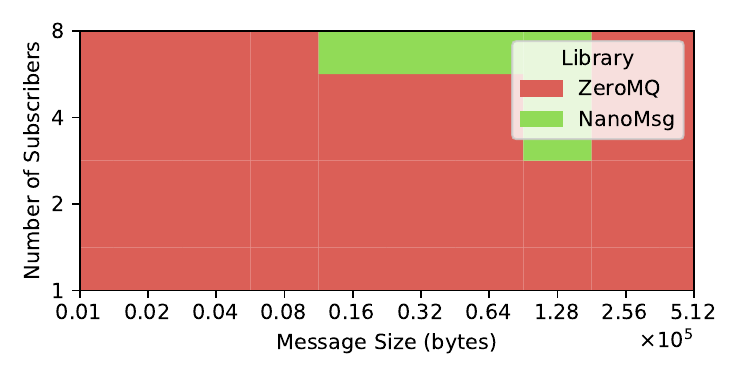}
        \caption{}\label{fig:opt-tcp-throughput}
    \end{subfigure}
    \hfill
    \begin{subfigure}[b]{0.30\linewidth}
        \centering
        \includegraphics[width=\linewidth]{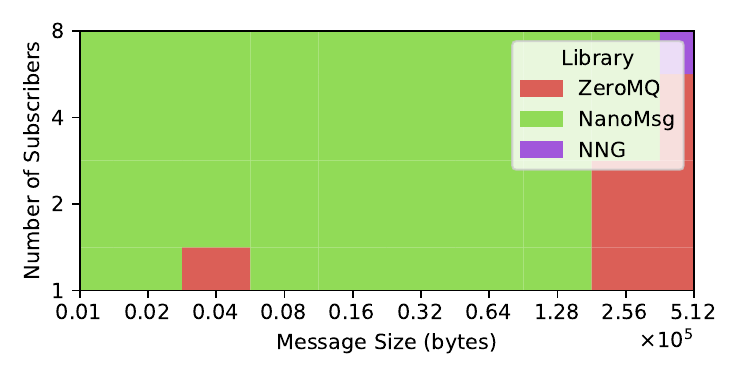}
        \caption{}\label{fig:opt-tcp-cpu}
    \end{subfigure}

    \caption{Regions of optimal performance for the three brokerless libraries evaluated.  
    Rows correspond to the transport mechanism --- In-Process (top), Inter-Process (middle), and TCP (bottom), while columns report, from left to right, latency, throughput, and median CPU usage results.  
    Within each panel, the shaded region indicates the library that attains the best result for every combination of message size and subscriber count explored with our tests.}
    \label{fig:optimality}
\end{figure*}

The In-Process communication scenario is shown in the top row of Figure~\ref{fig:optimality}. In the latency subplot (Figure~\ref{fig:opt-inproc-latency}), we observe that NanoMsg offers the lowest latency for small to medium message sizes (up to 32 or 64\,KB) across the various subscriber scenarios. However, ZeroMQ performs better at larger message sizes, confirming better outcomes in handling larger payloads. Regarding throughput (Figure~\ref{fig:opt-inproc-throughput}), ZeroMQ performs overall best, delivering higher throughput across all configurations of message sizes and subscribers, except some of the ones involving one subscriber or very small payloads (up to 4\,KB). For CPU efficiency (Figure~\ref{fig:opt-inproc-cpu}), again, NanoMsg often performs optimally up to the message size of 32\,KB, from where ZeroMQ usually proves to be more efficient.

The Inter-Process communication scenario is shown in Figure~\ref{fig:optimality}, middle row, where similar insights are observed. 
In the latency subplot (Figure~\ref{fig:opt-ipc-latency}), NanoMsg outperforms the other libraries for almost all message sizes and subscriber combinations, with ZeroMQ being preferable only at the very largest message sizes (256 and 512\,KB).

\begin{observation}\label{obs:lat-threshold}
\emph{In-Process and IPC transports present message size thresholds at which latency optimality shifts.}
\end{observation}

For throughput (Figure~\ref{fig:opt-ipc-throughput}), ZeroMQ is optimal across most configurations, except for regions involving small payloads and a single subscriber.
NanoMsg is the most CPU-efficient library in most of the Inter-Process configurations (Figure~\ref{fig:opt-ipc-cpu}), with a few exceptions where it is outperformed by ZeroMQ.
%

Finally, the TCP communication scenario is reported in the bottom row of Figure~\ref{fig:optimality}. The latency subplot (Figure~\ref{fig:opt-tcp-latency}) shows a more complex scenario, where NNG becomes relevant, as also mentioned in Section~\ref{sec:results_tcp}. The three libraries all have regions of optimal performance. In terms of throughput (Figure~\ref{fig:opt-tcp-throughput}), ZeroMQ performs well, delivering the highest throughput consistently for almost all scenarios tested. For CPU usage (Figure~\ref{fig:opt-tcp-cpu}), NanoMsg is again most efficient for nearly all the tested combinations, except at the largest message sizes (256 and 512\,KB), where NanoMsg or NNG slightly surpasses ZeroMQ in CPU efficiency.

\begin{observation}\label{obs:perf}
\emph{Across all transports and most configurations, ZeroMQ delivers the highest throughput, whereas NanoMsg achieves the best CPU efficiency.}
\end{observation}

%% file: related.tex
\section{Related Work}\label{sec:related}

Several studies examine messaging systems, mainly focusing on brokered architectures, with limited attention to brokerless systems.
\cite{celar2017state, yongguo2019message} review the state of the art of messaging systems, identifying brokerless libraries as emerging solutions.
Qualitative comparisons, highlighting ease of installation, documentation quality, and community support, are presented in \cite{patro2017comparative, fu2020fair, bertrand2021classification}. 
Quantitative evaluations of brokered systems are provided in \cite{yongguo2019message, fu2020fair, maharjan2023benchmarking, bertrand2021classification, de2019performance}.
\cite{fu2020fair} is particularly relevant because the authors develop their own testing framework to ensure a fair evaluation.
%
Although a few works \cite{patro2017comparative, barroso2016benchmarking, afanasev2017performance} include brokerless libraries among their candidates, they do not focus exclusively on such systems, instead evaluating them alongside brokered solutions or in specialized application contexts. 
In \cite{dworak2012new}, the authors evaluate Ice, ZeroMQ, and YAMI4 against specific workload and latency requirements in a PUB/SUB scenario, identifying ZeroMQ as the best-performing library, particularly due to its superior scalability with increasing numbers of subscribers. 
In \cite{barroso2016benchmarking}, ZeroMQ, NanoMsg, Asio, FairMQ, and O2 are tuned and benchmarked for a system transmitting terabits per second of data. The study analyzes network and memory throughput as a function of block size and system CPU usage, identifying Asio~\cite{asio} as the best solution. 
In \cite{afanasev2017performance}, the authors evaluate RabbitMQ, ActiveMQ, ZeroMQ, and NanoMsg for transmitting binary JSON data in distributed Computer Numerical Control (CNC) systems, concluding that NanoMsg is the best solution in terms of bandwidth, API flexibility, and POSIX compatibility.

To the best of our knowledge, no existing study includes NNG as a candidate, nor provides a dedicated, comprehensive benchmark specifically for brokerless messaging frameworks.
This gap in the literature is filled by our contribution.

%% file: conclusion.tex
\section{Conclusion and Future Work}\label{sec:conclude}

In this work, we presented a qualitative study and a systematic performance evaluation of brokerless messaging libraries.
First, we conducted a qualitative analysis based on licensing, documentation, and community support, which led us to select ZeroMQ, NanoMsg, and NNG as the most promising candidates.
Then, we developed and released a comprehensive benchmarking suite to evaluate latency, throughput, jitter, CPU usage, and memory consumption across different transports and workload conditions. Our framework is publicly available~\cite{suite}, allowing reproducibility and extensibility.

Our results suggest that selecting the 
most efficient
library largely depends on the specific workload and configuration, highlighting important trade-offs between the libraries;
%
specifically, ZeroMQ excels in throughput and offers the best performance at large payloads, whereas NanoMsg performs best for small message sizes and demonstrates consistent strength in CPU efficiency.
Except for some specific configurations, NNG performance is typically less competitive. However, it usually falls within the same order of magnitude. Moreover, NNG is the only library under active development.

In future work, we plan to extend our suite to include multi-machine communication, additional patterns, and libraries. 